\documentclass{ws-ijmpa}

\usepackage{epsfig}
\usepackage{graphicx}
\usepackage{dcolumn}
\usepackage{bm}

\newcommand{\beq}{\begin{equation}}
\newcommand{\eeq}{\end{equation}}
\newcommand{\bea}{\begin{eqnarray}}
\newcommand{\eea}{\end{eqnarray}}
\newcommand{\hf} {\frac{1}{2}}

\newcommand{\nn}{\nonumber\\}

\newcommand\eqn[1]{Eq.\,(\ref{#1})}

\newcommand\fig[1]{Fig.\,{\ref{#1}}}

\def\mr#1{{\mathrm{#1}}}

\def\t{\tilde}

\begin{document}

\markboth{S. Nagy, K. Sailer}
{Interplay of fixed points in scalar models}

\catchline{}{}{}{}{}

\title{Interplay of fixed points in scalar models}

\author{S. Nagy}

\address{Department of Theoretical Physics, University of Debrecen P.O. Box 5, H-4010 Debrecen, Hungary}
\address{MTA-DE Particle Physics Research Group, P.O.Box 105, H-4010 Debrecen, Hungary\\
nagys@dtp.atomki.hu}

\author{K. Sailer}

\address{Department of Theoretical Physics, University of Debrecen, P.O. Box 5, H-4010 Debrecen, Hungary\\
sailer@phys.unideb.hu}

\maketitle

\begin{history}
\received{Day Month Year}
\revised{Day Month Year}
\end{history}

\begin{abstract}
We performed the renormalization group analysis of scalar models
exhibiting spontaneous symmetry breaking. It is shown that an infrared
fixed point appears in the broken symmetric phase of the models, which
induces a dynamical scale, that can be identified with the correlation length.
This enables one to identify the type of the phase transition which shows
similarity to the one appearing in the crossover scale.
The critical exponent $\nu$ of the correlation length also
proved to be equal in the crossover and the infrared scaling regimes.
\end{abstract}

\ccode{PACS numbers: 11.10.Gh, 11.10.Hi, 05.10.Cc}

\section{Introduction}

The renormalization group (RG) method \cite{BTW,Polonyilect,Litim_2001,LitPaw,Pawlowski_2007,Delamottelect,Benitez_2009,Rostenlect}
is one of the best candidate to take into account
the degrees of freedom of a quantum field theoretical model systematically.
The method enables us to identify the fixed points of the models and the scaling of the couplings
in their vicinity which can give us the critical exponents of the corresponding fixed point
\cite{Tetradis_1994,Rosa,Nagy_2009}. The RG method is usually tested in 3-dimensional (3d) $O(N)$
scalar models, where a trivial Gaussian and a non-trivial Wilson-Fisher (WF) fixed point exist.
The calculation of the critical exponents of the latter fixed point plays the test ground
of all inventions in the RG method \cite{Tetradis_1994,Morris_1997,Morris_1998,Liao,Litim_2002,Delamotte1,Delamotte2,Litim_2010}.
The scaling of the correlation length $\xi$ defines the exponent $\nu$ as
\beq\label{nuis}
\xi \propto t^{-\nu},
\eeq
with the reduced temperature $t$. There is a second order, Ising type phase transition in the 3d
$O(1)$ model with exponent $\nu\approx 0.63$. The Kosterlitz-Thouless (KT) or infinite order phase
transition \cite{Ber,Kos,Grater,Kehrein,Gersdorff} is widely investigated too, furthermore it gives a great challenge
to recover the scaling of $\xi$ according to
\beq\label{nukt}
\log\xi \propto t^{-\nu}.
\eeq
A typical example for KT transition is presented by the sine-Gordon (SG) model
\cite{Nagy_2009,Amit,Coleman,ZJ,Nandori_2001,Nagy_2007,Pangon_2010,Nandori_2011,Nandori_2008,Kovacs_2011,Pangon_2012}
in two-dimensional (2d) Euclidean space, which belongs
to the same universality class as the 2d Coulomb gas and the 2d XY model. Furthermore the
SG model and its generalizations with compact variable have been thoroughly investigated
in the framework of integrable field theory \cite{Zam,Delfino,Mussardo}.

The fixed points and the typical scaling of $\xi$ in their vicinity
as in \eqn{nuis} for the 3d $O(N)$ and in \eqn{nukt} for the 2d SG models are located 
in the crossover region, i.e. between the ultraviolet (UV) and infrared (IR)
regimes. It has been argued \cite{Tetradis} that there exists an IR fixed point in the
3d $O(N)$ model at a finite momentum scale, which can be uncovered by a genuine rescaling of the
couplings around the singularity point of the RG evolution, i.e. where the flows stop.
Recently it has been obtained that there exists a similar IR fixed point in the SG model,
too \cite{Nagy_2009}. We note that both IR fixed points belong to the spontaneous symmetry
breaking phase of the models.

We show in this letter that this IR fixed point is accompanied by the appearing dynamical length
scale. It is defined by the scale of the RG evolution, where the flow equations become
singular, and can be identified with the reciprocal of
the correlation length. This provides us a method to determine the exponent $\nu$ beyond
the crossover scaling regime in the vicinity of the IR fixed point, which
accounts for the scaling of $\xi$. In the case of the 3d $O(1)$ model we obtain
numerically that the IR fixed point induced scaling gives the same exponent $\nu\approx 0.63$
that was obtained in the vicinity of the WF fixed point \cite{Liao,Litim_2002}.
The 2d SG model possesses a KT-type phase transition \cite{Nagy_2009}. There the reparametrization
of the flow equation suggested the existence of the IR fixed point for the Callan-Symanzik
type IR regulator \cite{Alex1,Alex2}. Here we show that the IR fixed point can be recovered
from a general type of regulator, furthermore the $\xi$, which is defined at the IR fixed point
scales as the one defined in the vicinity of the KT fixed point and,
similarly to the 3d $O(1)$ model, we get the same exponent $\nu\approx 1/2$
in both scaling regimes.

The method is also applicable when there is no crossover fixed point, e.g. in the bi-layer
SG model \cite{DeCol,Nandori_2006,Jentschura,Benfatto}, showing a greater flexibility and a more fundamental nature of the
method presented here.

\section{Evolution equation}

The Wetterich RG equation for the effective action is \cite{Wetterich}
\beq\label{WRG}
\dot\Gamma_k=\hf\mr{Tr}\frac{\dot R}{R+\Gamma''_k}
\eeq
where $^.=k\partial_k$, $^\prime=\partial/\partial\varphi$ and the trace Tr denotes the
integration over all momenta. \eqn{WRG} has been solved over the functional subspace defined by
the ansatz
\beq\label{eaans}
\Gamma_k = \int_x\left[\frac{Z_k}2 (\partial_\mu\varphi)^2+V_k\right],
\eeq
with the $O(1)$ symmetric or the periodic potential $V_k$ and the wavefunction renormalization
$Z_k$, which constant part is $z{=}Z_k(\varphi{=}0)$. The polynomial IR regulator has the form
$R = p^2(k^2/p^2)^b,$ with $b\ge 1$. \eqn{WRG} leads to the evolution equations for the couplings
\cite{BTW,Polonyilect,Litim_2001,LitPaw,Pawlowski_2007,Delamottelect,Benitez_2009,Rostenlect,Morris_1997,Morris_1998}.
The loop integral appearing in the RG equations should be
performed numerically when $b\ne 1$. The scale $k$ covers the momentum interval from the UV
cutoff $\Lambda$ to zero. Typically we set $\Lambda=1$. If we introduce $\bar k =\min(z p^2+R)$,
the RG evolution becomes singular at $k=k_f$ when
\beq\label{degen}
\left.\bar k^2 + V''_k(\varphi=0)\right|_{k=k_f}=0,
\eeq
where $\bar k^2 = b k^2[z/(b-1)]^{1-1/b}$, when $b=1$, then $\bar k=k$. The solution of this
equation defines the scale at which the action becomes degenerate. 

\section{Ising type transition}

\begin{figure}[ht] 
\begin{center} 
\epsfig{file=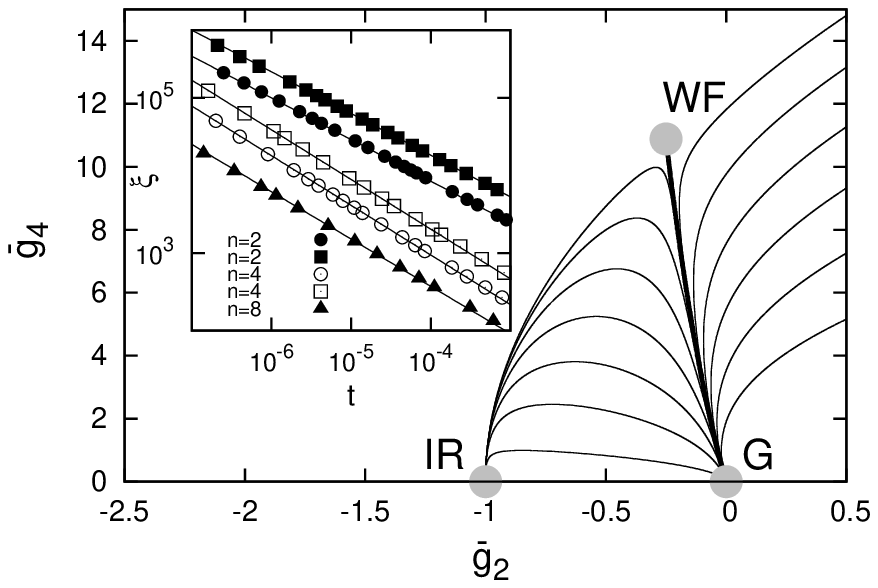,width=6cm,angle=-90}
\caption{\label{fig:is}
The phase structure of the 3d $O(1)$ model is presented. The thick curve shows the separatrix.
The trajectories tending to the right (left) correspond to the symmetric (symmetry broken) phase,
respectively. The fixed points are also shown.
The scaling of the correlation length as the function of the reduced temperature
$t$ is plotted in the inset. The points are obtained from the scaling around the IR fixed point,
while the solid lines represent the scaling around the WF fixed point, i.e. $\nu=0.53$ (LPA),
$\nu=0.64$ (LPA), and $\nu=0.62$ (including the flow of $Z_k$) for $n=2$, $n=4$, and $n=8$,
respectively. The curves are shifted for better visibility. The circle and square correspond
to $b=2,5$, respectively.}
\end{center}
\end{figure}

For the 3d $O(1)$ model the potential in \eqn{eaans} has the form
\beq\label{Vis}
V_k = \sum_{i=1}^n\frac{g_{2i}}{(2i)!}\varphi^{2i},
\eeq
with the couplings $g_{2i}$. One can take into account the evolution of
the wave function renormalization with similar ansatz for $Z_k$ as
\beq\label{Zis}
Z_k = z+\sum_{i=1}^n\frac{z_{2i}}{(2i)!}\varphi^{2i}.
\eeq
We also use the normalized couplings defined as $\bar x=x/\bar k^2$, where $x$ can be
$g_{2i},z,z_{2i}$. The evolution equations for the couplings $\bar g_2$ and $\bar g_4$ with
the choice $b=1$ are
\bea\label{g2g4}
\dot{\bar g}_2 &=& -2\bar g_2-\frac{\bar g_4}{8\pi(1+\bar g_2)^{1/2}},\nn
\dot{\bar g}_4 &=& -\bar g_4+\frac{3\bar g_4^2}{16\pi(1+\bar g_2)^{3/2}}.
\eea
The phase structure spanned by these couplings is plotted in \fig{fig:is}. There is a
phase transition in the model, and the phase space contains two fixed points, that can be
easily identified from Eqs. (\ref{g2g4}). There is a trivial UV Gaussian fixed point at the
origin. The linearization of the flow equations in its vicinity give two negative eigenvalues
$s_1=-1$ and $s_2=-2$ showing  that the UV fixed point is repulsive. The Wilson-Fisher fixed
point is a non-trivial one at the crossover scaling regime, and it can be found at
$\bar g_2^{*WF}=-1/4$ and $\bar g_4^{*WF}=\sqrt{12}\pi$. The corresponding eigenvalues coming
from the linearized flows are $s_1=4/3$ and $s_2=-2$. Since the critical exponent $\nu$
of the correlation length $\xi$ is identified as the negative reciprocal of the single
negative eigenvalue of the matrix coming from the linearization of the evolution equations,
the latter eigenvalue gives $\nu=-1/s_2=1/2$. The approximation of the
model with two couplings makes the problem a mean field type one.

However one can easily recognize from the phase structure, that in the broken symmetric phase
the trajectories tend to a single point at $\bar g_2^{*IR}=-1$ and $\bar g_4^{*IR}=0$.
It corresponds to the universal effective potential of the form $\t V_0=-\varphi^2/2$.
It suggests that this point is also a fixed point of the model, although this point
makes the flow equations in Eqs. (\ref{g2g4}) singular. By reparametrization of the couplings
according to $\omega=1+\bar g_2$, $\chi=\bar g_4/\omega$ and $\partial_\tau=\omega \partial_t$
one obtains
\bea\label{omch}
\partial_\tau \omega&=& 2\omega(1-\omega)-\frac{\chi \omega}{8\pi},\nn
\partial_\tau \chi &=& -\chi+\frac{\chi^2}{4\pi}.
\eea
The reparametrized flow equations enable one to recover the Gaussian ($\omega^{*G}=1$,
$\chi^{*G}=0$), and the WF ($\omega^{*WF}=3/4$, $\chi^{*WF}=4\pi$) fixed points, however
another one appears at $\omega^{*IR}=0$ and $\chi^{*IR}=4\pi$. The latter can be identified
with the IR fixed point where the trajectories of the broken symmetric phase meet.
The corresponding eigenvalues are $s_1=1$ and $s_2=3/2$ expressing the attractive nature
of the IR fixed point.

\begin{figure}[ht] 
\begin{center} 
\epsfig{file=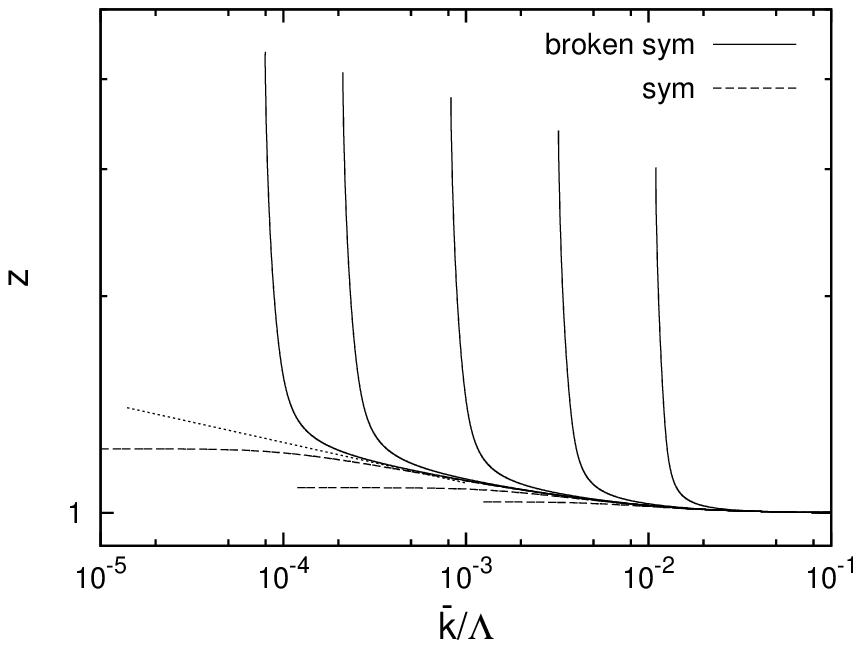,width=6cm,angle=-90}
\caption{\label{fig:isz}
The scaling of the $z$ for $n=8$ and $b=2$ for several UV values of $g_{4\Lambda}$.
The dotted line corresponds to the flow along the separatrix.
In the deep IR regime the flows blow up in the broken symmetric phase, while
they run into constant values in the symmetric one.
}
\end{center}
\end{figure}

If one considers the evolution of some couplings as the function of $k$, then one obtains
that they tend to infinity as $k\to 0$. However
if one plots their flows as the function of $\bar k$ then one gets that they
blow up at a certain scale $\bar k_f$. It is demonstrated in \fig{fig:isz},
where the coupling $z$ is plotted. The other couplings also have such singular behavior.
The flows do not run into real singularity as the function of $k$, and naturally
the effective potential keeps its convexity \cite{LPV}. The scale $\bar k$ can also be
considered as the momentum of the modes because it comes from transforming the Euclidean
propagator with the IR regulator into a dispersion relation-like form as in \eqn{degen}
In the broken symmetric phase a huge amount of soft modes appear, where the dispersion
relation gives infinitesimally small energies for the modes characterized by the momentum
$\bar k_f$. They create the appearing global condensate of size $1/\bar k_f$ in this phase
\cite{Wett_1991,Alex_1999,Braun}, which
is sometimes called as spinodal instability. This
argument makes the assumption plausible that the correlation length $\xi$ can be identified
with the reciprocal of the scale $\bar k_f$. We note that more precise results can be obtained
for the RG flows without Taylor expanding the potential and the wavefunction
renormalization \cite{Pangon_2011,TL}. In this treatment one typically obtains a marginal deep IR
evolution, however a significant change in the flow also appears at a certain scale $\bar k$,
and then this scale can be identified there by the reciprocal of the correlation length.

To get the exponent $\nu$, we fix the values of the UV couplings but $\bar g_{4\Lambda}$. At
a certain value of $\bar g_{4\Lambda}$ we determine the scale $\bar k_f$, where the singularity
appears during the flows. By fine tuning the value of $\bar g_{4\Lambda}$ to its critical UV
value $\bar g^*_{4\Lambda}$ one obtains smaller and smaller values for $\bar k_f$.
One can identify the reduced temperature as $t\sim \bar g_{4\Lambda}-\bar g^*_{4\Lambda}$
if the other UV values of the couplings are kept fixed.
\fig{fig:isz} demonstrates how the scale $\bar k_f$ of the singularity changes as $t\to 0$, the other
couplings show qualitatively similar pictures. The critical UV value $\bar g^*_{4\Lambda}$
can be got by the well-known trick, where one should fine tune its value on the log-log plot of
the $t,\xi$ plane till one obtains a straight line there. The negative slope of the line provides
us the exponent $\nu$.

We determined the exponent $\nu$ in the vicinity of the IR fixed point
by increasing the number $n$ in the lowest order of the gradient
expansion first, i.e. in the local potential approximation (LPA), when $Z_k=1$.
It was shown that $\nu$ at the WF fixed point is $\nu\approx 0.53$ ($\nu\approx 0.64$) for
$n{=}2$ ($n{=}4$), respectively \cite{Liao,Litim_2002}. We also investigated the scheme-dependence
of the results by choosing different values of $b$. The LPA results can be seen in
the inset of \fig{fig:is} for $n=2$ and $n=4$ couplings, which shows the coincidence of the
exponents calculated from the data around the WF and the IR points, and demonstrates that one
can determine the value $\nu$ in the IR, too.
We numerically determined the exponent $\nu$ for $n=8$ couplings beyond LPA from data around the
WF and the IR points, which, similarly to the LPA results, shows high coincidence, as is
demonstrated in the inset of \fig{fig:is}. The numerical results also show that the wavefunction
renormalization constant $z$ blows up in the vicinity of the degeneracy as the function of
$\bar k$, see \fig{fig:isz}.
The blowup of $z$ appears at $\bar k_f$ in the broken symmetric phase with all of the other
couplings, while in the symmetric phase $z$ goes to a constant value, giving LPA evolution in
the IR. The flow of $z$ denoted by dotted line in \fig{fig:isz} correspond to the
flow along the separatrix in \fig{fig:is}. Its slope gives the known, tiny anomalous dimension
$\eta$ $=-d\log z/d\log k\approx 0.05$ belonging to the WF fixed point. The flows in the vicinity
of the WF fixed point pick up the effects coming from the fluctuations affected by the WF fixed
point and bring them to the IR fixed point. There the anomalous dimension is extremely high
(and scheme dependent), but the exponent $\nu$ must not change, since it characterizes the global
condensate of the broken phase throughout the flow from the UV till the IR.

\section{Periodic model}

\begin{figure}[ht] 
\begin{center} 
\epsfig{file=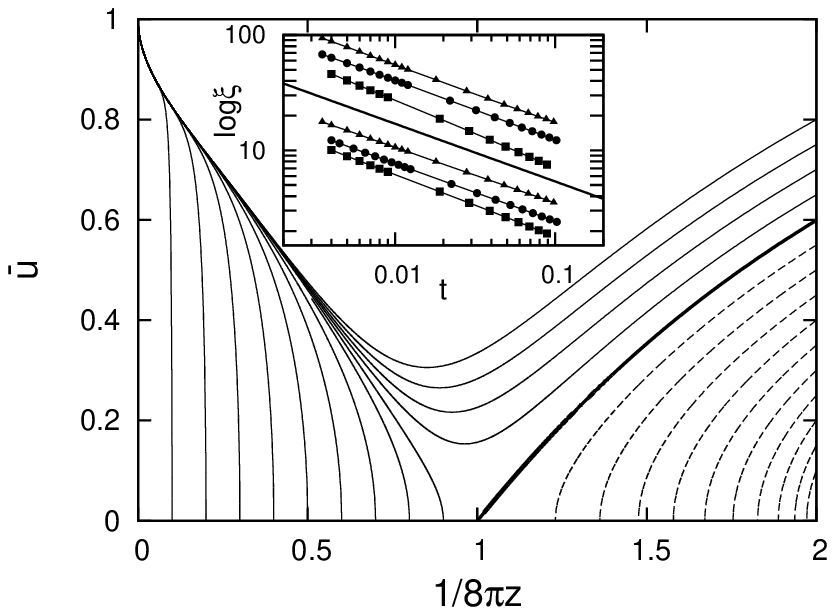,width=6cm,angle=-90}
\caption{\label{fig:sg}
Phase diagram of the SG model, with $b=5$. The dashed (solid) lines represent
the trajectories belonging to the (broken) symmetric phase, respectively.
The wide line denotes the separatrix between the phases. The inset shows the
scaling of the correlation length as the function of the reduced temperature $t$. The curves
are shifted for better visibility. The lower (upper) set of lines corresponds to the IR (KT) fixed
point. The triangle, circle and square correspond to $b=2,5,10$, respectively.
In the middle a straight line with the slope $-1/2$ is drawn to guide the eye.}
\end{center}
\end{figure}

The 2d SG model is defined via the effective action in \eqn{eaans} with the potential of the form
\beq\label{Vsg}
V_k = u \cos\varphi.
\eeq
The higher harmonics of the SG model are neglected. They correspond to vortices with higher
vorticity of the equivalent gas of topological excitations. It is known that only the fundamental
mode plays a significant role in the determination of the thermodynamic properties of
the model, while effects corresponding to higher vorticity are negligible. We also note
that the fundamental mode can recover the phase structure of the model including the
KT transition point \cite{Amit} and the IR behavior \cite{Nagy_2009}.
Furthermore the wavefunction renormalization constant $z$ can account for the KT
transition \cite{Nagy_2009,Grater,Kehrein,Gersdorff}, therefore we omit further terms in $Z_k$.
In the LPA approximation, the RG treatment of the SG model in the IR shows two phases separated by
the Coleman point at the value of parameter $z^*=1/8\pi$ \cite{Coleman}. The flows with
$z>1/8\pi$ ($z<1/8\pi$) correspond to evolutions in the broken symmetric (symmetric) phase,
respectively. The dynamical momentum scale turns up in the broken symmetric phase, where the
evolution of the normalized coupling $\bar u=u/\bar k^2$ becomes marginal
\cite{ZJ,Nandori_2001,Nagy_2007,Pangon_2010,Nandori_2011}. Identifying
this scale as the reciprocal of the correlation length we obtain $\nu=1$. The same universal
effective potential $\t V_0=-\varphi^2/2$ appears when $z\to\infty$ as in the case of the 3d $O(1)$
model. The Coleman point becomes the KT transition point \cite{Nagy_2009,Amit} if $Z_k$ evolves.
Furthermore an additional IR fixed point turns up that can be transformed to the unique point
at $\bar u=1$ and $1/z\to 0$ when the normalized coupling $\bar u$ is made of use
\cite{Nagy_2009}. Then any choice of $b$ gives qualitatively similar phase diagrams. We plotted
the case $b=5$ in \fig{fig:sg}.
The normalized coupling $\bar u$ tends to 1 for every value of $b$. It shows that the degenerate
potential (which satisfies \eqn{degen}) occurs in the IR limit of the broken symmetric
phase independently of the RG scheme. This reflects the serious limitation of the LPA
results. In the symmetric phase the evolution of $z$ is negligible
giving the same evolution as was obtained in LPA with the line of fixed points.

\begin{figure}[ht]
\begin{center}
\epsfig{file=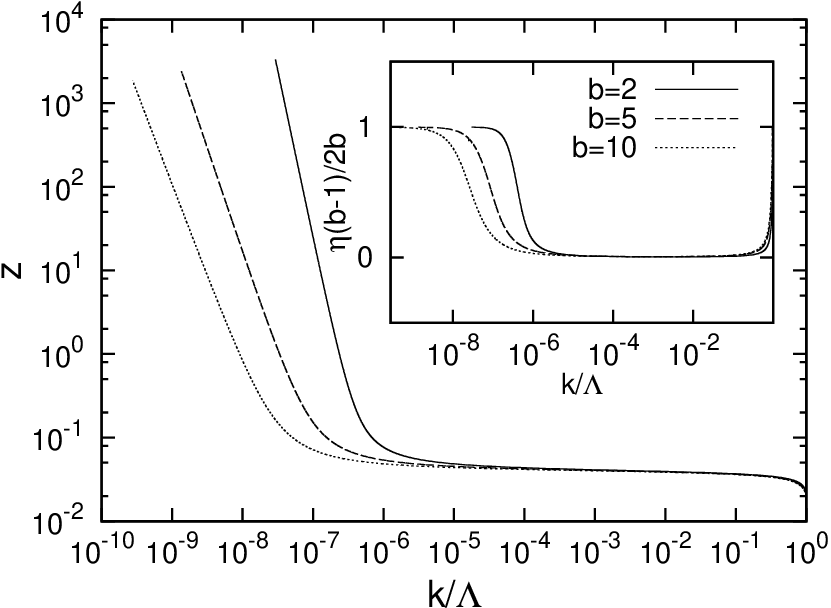,width=6cm,angle=-90}
\caption{\label{fig:eta}
The evolution of $z$ for different values of $b$. The inset demonstrates that $\eta=2b/(b-1)$
in the deep IR region.}
\end{center}
\end{figure}

One can easily show that the critical exponent $\eta_v$ characterizing the vortex-vortex
correlation function \cite{Kosterlitz} is $\eta_v=1/4$ independently of the parameter $b$
\cite{Nagy_2009}. However the anomalous dimension being characteristic for the divergence of the
correlation function of the field variables gives $\eta=0$ in the vicinity of the KT point.
In the deep IR scaling region the situation changes significantly, there new scaling laws appear.
\fig{fig:eta} shows that around the KT point (at about $k/\lambda\sim 10^{-4}$) $z$ 
is practically constant, giving $\eta=0$, while in the IR region $z$ diverges
with the exponent $\eta=2b/(b-1)$, giving scheme-dependent $\eta$ values in the IR.
It implies that one cannot avoid the evolution of $z$ by including a phase factor
$z=k^{\eta}$ in the LPA evolution equations.
The scale $\bar k$ makes the evolution of $u$ and $z$ qualitatively scheme-independent.
The value of the wavefunction renormalization $z$ blows up at a certain minimal value of $\bar k$
in the broken symmetric phase, similarly to the 3d $O(1)$ model. Likewise, the IR fixed point
of the broken symmetric phase is reached at $k=0$, but a new scaling behavior appears at
some finite dynamical scale $\bar k_f$.

The initial UV value of $z_\Lambda$ can be identified with the square of the temperature,
thus its distance from the separatrix (for fixed $u_\Lambda$) gives the reduced temperature $t$.
As $t\to 0$ the correlation length increases as in the case of the 3d $O(1)$ model.
Our numerical results are shown in the inset of \fig{fig:sg}. There are two types of correlation
lengths, one is defined as usual, namely at around the KT turning point of the coupling $\bar u$.
Another one is identified as $\xi=1/\bar k_f$ in the neighborhood of the IR fixed point. 
It can be seen from the inset of \fig{fig:sg} that the scaling of $\xi$ shows an infinite order
phase transition, for all schemes with the exponent $\nu\approx 0.5$.

\section{Conclusions}

The renormalization group treatment is performed for the 3d $O(1)$ and the 2d sine-Gordon
models, and it was shown that their broken symmetric phase possesses an IR fixed point, that
generates a new scaling regime there. The critical exponent $\nu$ of the correlation length $\xi$
is determined in its vicinity and it was found that it equals to that one
obtained in the crossover regime at the WF and KT fixed points, respectively. The IR
fixed point is the signal of spontaneous symmetry breaking in both the polynomial and the
periodic models, while the  WF and KT  points are crossover fixed points, although they are closely
related to the IR one. The dependence of the correlation length $\xi$ on the reduced temperature
has already been picked up by the RG trajectory in the crossover regime and is carried by it
to the IR fixed point. In that sense there is an interplay between the IR and crossover points.
The IR fixed point should recover all the information
on the correlation length, implying the information on the type of the phase transition,
since it should characterize the global condensate appearing in the broken symmetric phase.
The value of the anomalous dimension $\eta$ differs when calculated in the crossover and
in the IR regimes. The models investigated by us gave large anomalous dimension in the IR limit.
This might reflect the loss of locality in the low energy broken symmetric phase due to the
appearing global condensate and suggests that one should take into account higher order terms
in the gradient expansion.

On the one hand, the determination of the correlation length $\xi$ from the scaling in the
deep IR regime is also powerful and can provide us the value of $\nu$
when we have no crossover scalings. In the case of the bi-layer SG (LSG) model \cite{DeCol,Nandori_2006,Jentschura,Benfatto}
one can easily show that the model has no crossover fixed point because of the evolution of the
interlayer coupling. However, the detailed RG investigations performed by us have shown
that there is an attractive IR fixed point of the LSG model with the appearing dynamical
scale $\bar k_f$ as in the 2d SG model, and gives the same KT type phase transition
with exponent $\nu\approx 0.57$ \cite{Nagy_per}, proving the infinite nature of the phase
transition there.

On the other hand, this method makes easy to determine numerically the value of $\nu$,
since one should only fine tune the UV value of a single coupling. As opposed to the general
treatment there is no need to have information on the place of the fixed point, which is
difficult to find numerically, especially if the dimension of the phase space is large.
Finally we notice, that this method is capable of characterizing the IR fixed point
and can uncover the IR physics of scalar models by taking into account a few terms in the
expansion of the potential and the wavefunction renormalization, although there are much
reliable treatments available nowadays, based on considering the Wetterich equation without any
expansion for them. However there are several cases where the more involved treatments
cannot be applied, e.g. in the case of quantum gravity
\cite{Reuter,Bonanno,Litim_2004,ReuSau,Donkin,NagyIR}, where the flow equations of
only a few couplings can be obtained. By using this method one can easily describe its IR physics,
and find the existing IR fixed point there.

\section*{Acknowledgments}

This research was supported by the European Union and the State of Hungary,
co-financed by the European Social Fund in the framework of T\'AMOP 4.2.4. A/2-11-1-2012-0001 National Excellence Program.
(Author: S\'andor Nagy)

\end{document}